\documentclass[a4paper,fleqn,usenatbib]{mnras}

\usepackage{newtxtext,newtxmath}

\usepackage[T1]{fontenc}
\usepackage{ae,aecompl}

\usepackage{graphicx}	
\usepackage{amsmath}	
\usepackage{amssymb}	

\newcommand{\rev}{\textcolor{black}}
\newcommand{\mpo}{\textcolor{black}}

\title[PIC simulation of jet evolution]{Rapid Particle Acceleration due to Recollimation Shocks and  Turbulent Magnetic Fields in Injected  
Jets with Helical Magnetic Fields}

\author[K. Nishikawa et al.]{
Kenichi Nishikawa,$^{1}$\thanks{E-mail: kenichi.nishikawa@aamu.edu(KN)}
Yosuke Mizuno,$^{2}$
Jose L. G\'omez,$^{3}$
Ioana Du\c{t}an,$^{4}$
\newauthor Jacek Niemiec,$^{5}$
Oleh Kobzar,$^{5}$
Nicholas MacDonald,$^{6}$
Athina Meli,$^{7,8}$
\newauthor Martin Pohl$^{9}$
and Kouichi Hirotani$^{10}$
\\
$^{1}$Department of Physics, Chemistry and Mathematics, Alabama A\&M University,
Normal, AL 35762, USA\\
$^{2}$Institute for Theoretical Physics, Goethe University, D-60438 Frankfurt am Main, Germany\\
$^{3}$Instituto de Astrof\'isica de Andaluc\'ia, CSIC, Apartado 3004, 18080 Granada, Spain\\
$^{4}$Institute of Space Science, Atomi\c{s}tilor 409, RO-077125 Bucharest-M\u{a}gurele, Romania\ \\
$^{5}$ Institute of Nuclear Physics Polish Academy of Sciences, PL-31342 Krakow, Poland\\
$^{6}$Max-Planck-Institut f\"{u}r Radioastronomie, Auf dem H\"{u}gel 69, D-53121 Bonn, Germany\\
$^{7}$Space Sciences \& Technologies for Astrophysics Research (STAR) Institute Universite de Liege, Sart Tilman,
4000 Li{\'{e}}ge, Belgium\\
$^{8}$Department of Physics and Astronomy, University of Gent, 9000 Gent, Belgium\\
$^{9}$Institute of Physics and Astronomy, University of Potsdam, 14476 Potsdam-Golm,
and DESY, Platanenallee 6, 15738 Zeuthen, \\Germany\\
$^{10}$Academia Sinica, Institute of Astronomy and Astrophysics (ASIAA), PO Box 23-141, Taipei, Taiwan
}

\date{Accepted 2020 February 5. Received 2020 January 17; in original form 2019 October 15}

\pubyear{2015}

\begin{document}
\label{firstpage}
\pagerange{\pageref{firstpage}--\pageref{lastpage}}
\maketitle

\begin{abstract}
One of the key questions in the study of relativistic jets is how magnetic reconnection occurs and whether it can effectively accelerate \mpo{electrons in}
the jet. \mpo{We performed 3D particle-in-cell (PIC) simulations of a relativistic electron-proton jet of relatively large radius that carries a helical magnetic field. We focussed our investigation on the 
interaction between the jet and the ambient plasma and explore how the helical magnetic field affects} the excitation of kinetic instabilities such as 
the Weibel instability (WI), the kinetic Kelvin-Helmholtz instability (kKHI), and the mushroom instability (MI). In our simulations these kinetic 
instabilities are indeed excited, and particles are accelerated. \mpo{At the linear stage} we observe recollimation shocks near the center of the jet.  
As the electron-proton jet evolves \mpo{into the deep nonlinear stage}, the helical magnetic field becomes untangled due to reconnection-like phenomena, and electrons are \mpo{repeatedly accelerated as they encounter} magnetic-reconnection events in the turbulent magnetic field.
\end{abstract}




\section{Introduction}

Magnetic reconnection is ubiquitous in solar and magnetospheric plasmas, and it has been proposed
that it is also an important mechanism for particle acceleration in Active Galactic Nuclei (AGN) and gamma-ray burst jets
\citep{Drenkhahn2002,Pino05,Pino10,uzdensky2011,granot2011,granot2012,mcKinney2012,zhang2011,gianninos2009,
gianninos2010,gianninos2011,   
komissarov2012,sironi2015,Pino18,Kadowaki18,Kadowaki19,Christie19,Fowler19}.
To study the fundamental physics of magnetic reconnection, numerous
particle-in-cell (PIC) simulations have been performed that use the Harris model in a slab geometry
\cite[e.g.,][]{daughton2011,wendel2013,karimabadi14,zenitani2005,zenitani2008,zenitani2011,zenitani2013,oka2008,fujimoto2011,
kagan2013,sironi2011,sironi2014,guo2015,guo2016a,guo2016b}. Generally, studies in the slab configuration \mpo{show} significant particle 
acceleration. However, this configuration \mpo{does not apply to astrophysical jets, since the magnetic field appears to have a helical morphology} close to the jet collimation/launching point \cite[e.g.,][]{tchekhovskoy2015,hawley2015,gabuzda2019}.

\mpo{In jets MHD instabilities such as \rev{Kelvin-Helmholtz} (KHI) and \rev{the kink instability} (KI) can operate} \cite[e.g.,][]{birkshaw84,birkshaw96,stone1994,stone2000}.
Recently, these MHD instabilities have been revisited using PIC simulations including kinetic processes \cite[e.g.,][]{sironi2013,alves15,Ardaneh16,nishikawa2019}.

The interaction of relativistic jets with the plasma environment generates relativistic shocks that are mediated by the Weibel instability (WI) 
and accelerate particles. \mpo{At the velocity-shear boundary between the jet and the ambient medium} magnetic turbulence is generated by the kinetic Kelvin-Helmholtz (kKHI) and 
the mushroom (MI) instability. 
\mpo{Recent PIC simulations explored the WI, kKHI and MI  in slab models of jets and later focussed on the evolution of cylindrical jets}
in helical magnetic-field geometry \cite[e.g.,][]{sironi2013,alves15,Ardaneh16,nishikawa2019}.
To achieve a complete understanding 
of the physics within the jet, 
global three-dimensional (3D) modeling \rev{needs to be performed} that enables investigation of the combined shock/shear \rev{layer}
processes and includes kinetic effects. Our first PIC simulations of global jets were performed for unmagnetized plasmas \citep{nishikawa2016a}. 
With this work we extend these studies to jets containing helical magnetic field.

One of the key questions we want to answer is how the helical magnetic field affects the growth of the kKHI, MI, and WI, and how and where 
in the jet structure particles are accelerated. In the latter respect, \mpo{we are particular interested in magnetic reconnection and} its ability to aid 
in the rapid merging and breaking of the helical magnetic field 
carried by relativistic jets.  Relativistic magnetohydrodynamic (RMHD) simulations demonstrate that jets with helical field develop kink 
instabilities (KI) \cite[e.g.,][]{mizuno2014,singh2016,barniol2017,Pino18,Kadowaki18,Kadowaki19}; similar structures were found 
in PIC simulations \cite[see, e.g.,][]{nishikawa2019}.
It should be noted that MHD instabilities such as KHI and KI  in jets have been investigated extensively \citep[e.g.,][]{birkshaw84,stone1994,stone2000,
hawley2015}.
PIC simulations of a single flux rope modeling the jet that undergoes internal KI showed signatures of secondary magnetic reconnection 
\citep{markidis14}. 

Recently, it was demonstrated that the development of the KI in relativistic strongly magnetized jets with helical magnetic field 
leads to the formation of highly tangled magnetic field and a large-scale inductive electric field promoting the rapid energization 
of particles through curvature drift acceleration \citep{alves18,alves19}. As initial condition these simulations assumed 
helical magnetic field in the jet frame supported by \mpo{counter-streaming} electrons and positrons (ions). \mpo{There is neither velocity shear nor a jet head in their simulation, and so} no velocity-shear instability such as kKHI and MI can be excited.

In this work we present results of our new study of relativistic jets containing helical magnetic field, which exhibit the nonlinear evolution of 
kinetic instabilities, magnetic reconnection, and the associated particle acceleration. We focus on electron-scale phenomena and set the jet size sufficient to accommodate the relevant electron 
kinetic instabilities that are not included in MHD simulations. The kinetic instabilities grow to the nonlinear stage, which is demonstrated 
by the disappearance of the helical magnetic field. Although our simulation results are insufficient to model the large-scale plasma flows of 
macroscopic parsec-scale jets, they explore relevant kinetic-scale physics within relativistic jet plasma.

\section{PIC Simulation Setup of a Jet with Helical Magnetic Field Structure}
\label{sec:2.1.}
\vspace*{-0.0cm}

In our simulations a cylindrical jet containing  helical magnetic field \mpo{and propagating in the $x$-direction is injected into an ambient plasma at rest,} 
as is shown schematically in  Fig. 6a in \cite{nishikawa2019}. The structure of the helical magnetic field is implemented in the same way as in the RMHD simulations 
by \citet{mizuno2014}, where a force-free expression of the field at the jet orifice is used.
The magnetic field is thus not generated self-consistently by a rotating black hole, as in GRMHD simulations of jet formation. 
\mpo{The initial helical magnetic field has the same form as} in Equations (1) and (2) of \citet{mizuno2014},
\begin{eqnarray}
B_{x} = \frac{B_{0}}{[1 + (r/a)^2]}, \,  \, \, \, \, \,  B_{\phi} =  \frac{(r/a)B_{0}}{[1 + (r/a)^2]},
\label{hmfcp}
\end{eqnarray}
where $r$ is the radial coordinate in cylindrical geometry, $B_0$ parametrizes the magnetic-field strength, and $a$ is the characteristic 
radius of the magnetic field \citep{nishikawa2019}. 
Note that for a constant magnetic pitch, $\alpha=1$, the toroidal field component has a maximum at $r=a$. 
The toroidal component of the magnetic field in the jet is created by an electric current, $+J_{x}(y, z)$, in the positive $x$-direction. 
In Cartesian coordinates (used in our simulation) the corresponding $B_y$ and $B_z$ field components are defined as:
\begin{eqnarray}
B_{y}(y, z) =  \frac{((z-z_{\rm jc})/a)B_{0}}{[1 + (r/a)^2]}, \, \, \,\,  \,  \,
B_{z}(y, z) =  -\frac{((y-y_{\rm jc})/a)B_{0}}{[1 + (r/a)^2]}.
\label{hmfcar}
\end{eqnarray}
Here,  the center of the jet is located at $(y_{\rm jc},\, z_{\rm jc})$, and $r = \sqrt{(y-y_{\rm jc})^2+(z-z_{\rm jc})^2}$. Equation~\ref{hmfcar} 
defines the helicity of the magnetic field that has a left-handed polarity for positive $B_{0}$. At the jet orifice, the helical magnetic field is 
computed without motional electric fields. This corresponds to magnetic-field generation by jet particles moving \rev{in} the $+x$-direction.

The simulated jet has a radius $r_{\rm jet}$  and is assumed to propagate in an initially unmagnetized ambient medium. For the fields 
external to the jet we use a damping function, $\exp{[-(r-r_{\rm jet})^2/b]} \, (r\geq r_{\rm jet})$, that multiplies the expressions in 
Equation~\ref{hmfcar} with the tapering parameter \mpo{$b=200\Delta$, where $\Delta $ is the grid scale}. We further assume a characteristic radius $a =  0.25*r_{\rm jet}$. 
The profiles of the resulting helical magnetic field components are shown in Figure~6b in \cite{nishikawa2019}. The toroidal magnetic field 
\mpo{vanishes} at the center of the jet (red line \rev{in}  Fig. 6b of \cite{nishikawa2019}). Note, that the simulation 
setup adopted in this work has been used in our preliminary studies of helical jets \citep{nishikawa2016b,nishikawa2017,nishikawa2019} 
with the modifications $B_0 = 0.1$ and $r_{\rm jet} =100$.


\begin{figure*}
	\includegraphics[scale=.98]{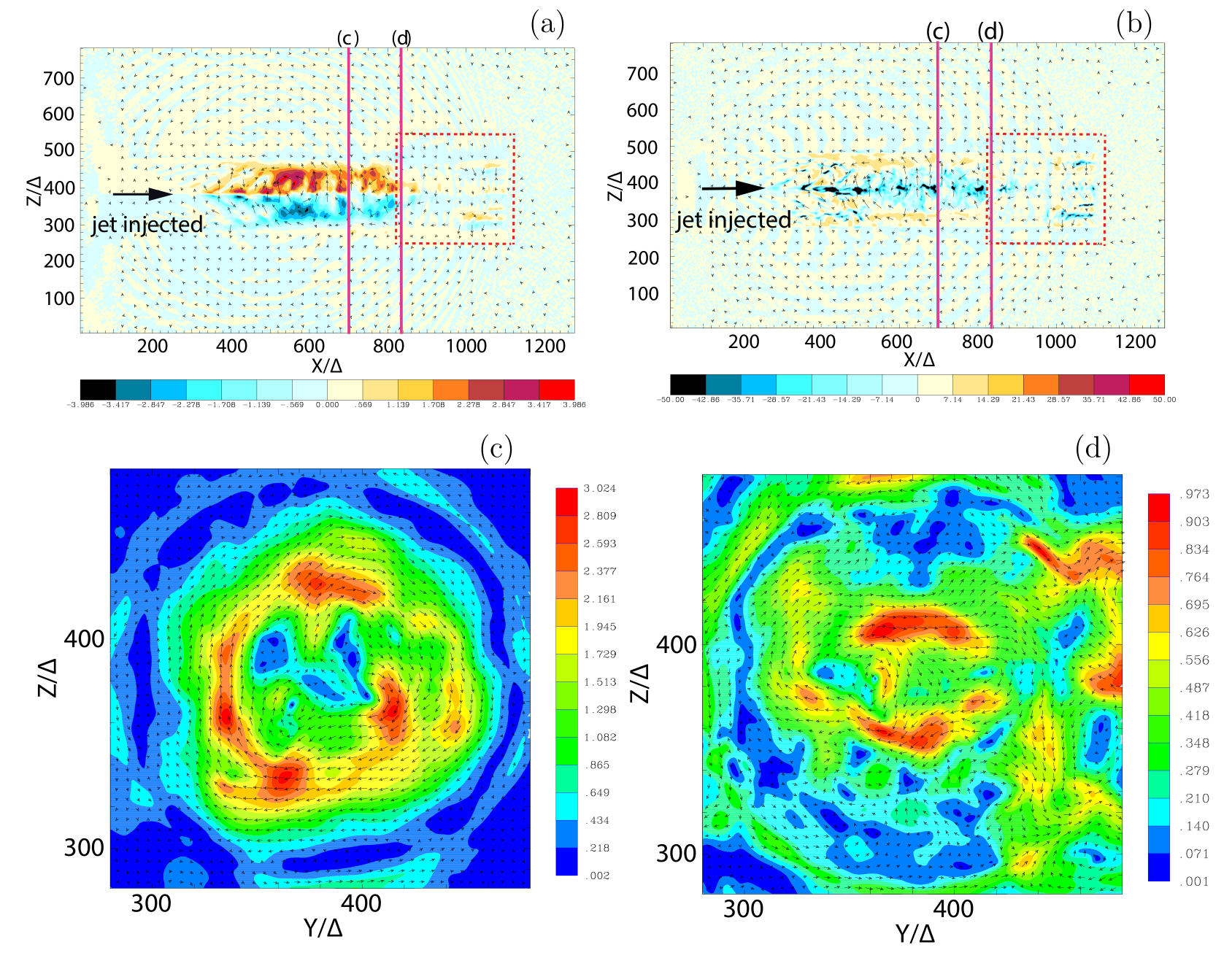}
\caption{Upper panels: (a) the $y$-component of the magnetic field, $B_{\rm y}$, with $x$-$z$ electric field depicted by arrows, and (b) 
the $x$-component of the electron current density, $J_{\rm x}$, with the $x$-$z$ magnetic field as arrows, both in the $x-z$ plane 
at $t =1000\, \omega_{\rm pe}^{-1} $.  The lower panels show the total magnetic-field strength in the $y-z$ plane at $x/\Delta=700$ 
(c) and $x/\Delta=835$ (d). The arrows indicate the magnetic field ($B_{\rm y}, B_{\rm z}$).}
\label{JxB}
\end{figure*}

The jet head \mpo{assumed here has a flat-density top-hat shape which is a strong simplification of the true} structure of the jet-formation region \cite[e.g.,][]{broderick2009,moscibrodzka2017}. A more realistic jet structure (with, e.g., a Gaussian-shaped head) will be implemented in future studies.

The numerical code we use is a modified version of the relativistic electromagnetic PIC code TRISTAN \citep{tristan} with MPI-based 
parallelization \citep{niemiec_2008}. The simulations are performed in Cartesian coordinates on a numerical grid of size 
\mbox{$(L_{x}, L_{y}, L_{z}) = (1285\Delta, 789\Delta, 789\Delta)$}. We use open boundaries  at the
$x/\Delta=0$ and $x/\Delta=1285$ surfaces and impose periodic boundary conditions in the transverse directions. Since the jet \rev{is} 
located at the center of \rev{the numerical box} far from the boundaries and the simulation is rather short, 
\rev{there are no visible effect of the} periodic boundary conditions \rev{on the system.} 

The jet radius is $r_{\rm jet} = 100\Delta$.  The jet is injected at $x = 100\Delta$ in the center of the $y-z$ plane. The large computational 
box allows us to follow the jet evolution \mpo{long enough to permit the investigation of the} strongly nonlinear stage. The longitudinal 
box size, $L_{x}$, and the simulation time, $t_{\rm max}=1000\omega_{\rm pe}^{-1}$, are a factor of two  larger than  in our previous 
studies \citep{nishikawa2016b,nishikawa2017,nishikawa2019}, in which jets with radii $r_{\rm jet} = 20, 40, 80, 120\Delta$ were investigated.

We know that jets of different plasma composition exhibit distinct dynamical behavior \mpo{that manifests itself in the morphologies of 
the jet and its emission characteristics} \citep{nishikawa2016a,nishikawa2016b,nishikawa2017,nishikawa2019}.
In this report \mpo{we discuss only electron-proton plasma in} both the jet and the ambient medium.
The initial \mpo{particle number per cell is} $n_{\rm jt}= 8$ and  $n_{\rm am} = 12$,  
respectively, for the jet and ambient plasma. The electron skin depth is $\lambda_{\rm se} =  c/\omega_{\rm pe} = 10.0\Delta$, 
where $c$ is the speed of light, $\omega_{\rm pe} = (e^{2}n_{\rm am}/\epsilon_0 m_{\rm e})^{1/2}$ is the electron plasma frequency;
the electron Debye length of ambient electrons is $\lambda_{\rm D}=0.5\Delta$. The \mpo{thermal speed of jet electrons is}
$v_{\rm jt,th,e} = 0.014c$ in the jet reference frame; in the ambient plasma it is $v_{\rm am,th,e} = 0.03c$. \mpo{We assume temperature equilibration, and so the thermal speed of ions is smaller than that of electrons by the factor $(m_{\rm p}/m_{\rm e})^{1/2}$. We use the} realistic proton-to-electron mass ratio $m_{\rm p}/m_{\rm e}=1836$. 
The jet plasma is initially weakly magnetized, and the magnetic field amplitude parameter assumed, $B_{0}=0.1c$, corresponds 
to plasma magnetization \mbox{$\sigma = B^2/n_{\rm e}m_{\rm e}\gamma_{\rm jt}c^{2} =2.8\times 10^{-3}$}. The jet Lorentz 
factor \rev{is} $\gamma_{\rm jt}=15$.

\section{Structure of Helically Magnetized Jets}
\label{sec:2.2.}

Figure~\ref{JxB} shows cross-sections through the center of the jet at time $t =1000\, \omega_{\rm pe}^{-1}$ with (a) the $y$-component of the 
magnetic field,  $B_{\rm y}$, with the $x$-$z$ electric field as arrows, and (b), the $x$-component of the electron current density, $J_{\rm x}$, 
with the $x$-$z$ magnetic field depicted as arrows. The jet propagates from the left to right. 
\mpo{Very strong helical magnetic field in the jet is evident at $400\lesssim x/\Delta \lesssim 830$. The amplitude is with $B/B_0\approx 40$ (Fig.~\ref{JxB}a) much larger than that of the initial field.} As in the unmagnetized case \citep{nishikawa2016a}, this field results from MI and kKHI. 
However, in the presence of the initial helical magnetic field the growth rate of the transverse MI mode is reduced, and the field structure is strongly 
modulated by longitudinal kKHI wave modes as shown by the bunched  $B_{\rm y}$ field (Fig.~\ref{JxB}a). This causes multiple collimations 
along the jet \mpo{that are caused by pinching of the} jet electrons toward the center of the jet, as visible in the electron current density (Fig.~\ref{JxB}b). 
It should be noted that \mpo{in Fig \ref{JxB}b the color scale for $J_{\rm x}$ does not extend beyond $J_{\rm x} = -50$, as we intend} to show the weak positive (return) 
current. \mpo{The collimations become weaker further along the jet and eventually disappear} at  $x/\Delta\gtrsim 830$. 
At this point $B_{\rm y}$ is weak. This demonstrates that the nonlinear saturation of the MI \mpo{leads to the }
dissipation of the helical magnetic field.

\mpo{We selected possible reconnection sites in Figures~\ref{JxB}a-b, indicated by the two red lines at $x/\Delta=700$ and $x/\Delta=835$, and show the magnetic-field structure in the $y-z$ plane in figure panels}~\ref{JxB}c and \ref{JxB}d, respectively.At $x/\Delta=700$ clockwise-circular magnetic field is split near the jet into a number of magnetic structures, which \mpo{demonstrate} 
the growth of MI. They are surrounded by \mpo{field of opposite polarity that is produced by the proton 
current} framing the jet boundary \citep[see][]{nishikawa2016a}. \mpo{The magnetic field at $x=835\Delta$ is strongly turbulent; its helical structure is distorted and reorganised into multiple magnetic islands, which reflect the nonlinear stage}
of MI and kKHI. The  magnetic islands interact with each other, providing conditions for magnetic reconnection. 
In our 3D geometry reconnection does not occur at a simple X-point as in 2D slab geometry. Instead, reconnection sites can be identified 
with regions of weak magnetic field surrounded by oppositely directed magnetic field lines. 
An example of a possible location of reconnection can be found at $(y/\Delta, z/\Delta)=(380, 340)$, where the total magnetic field 
becomes minimal (the null point, Fig.~\ref{JxB}d). The evolution of the magnetic field at different locations in the jet 
($600 < x/\Delta < 1100$) is shown in the supplemental movie. Note that the filamentary structure at the jet head (Fig.~\ref{JxB}a-b) is 
formed by the electron WI. One can see in the movie that nonlinear evolution of the filaments also leads to the appearance of the magnetic 
structures that are prone to reconnection.


\begin{figure}
\includegraphics[scale=0.43,angle=0]{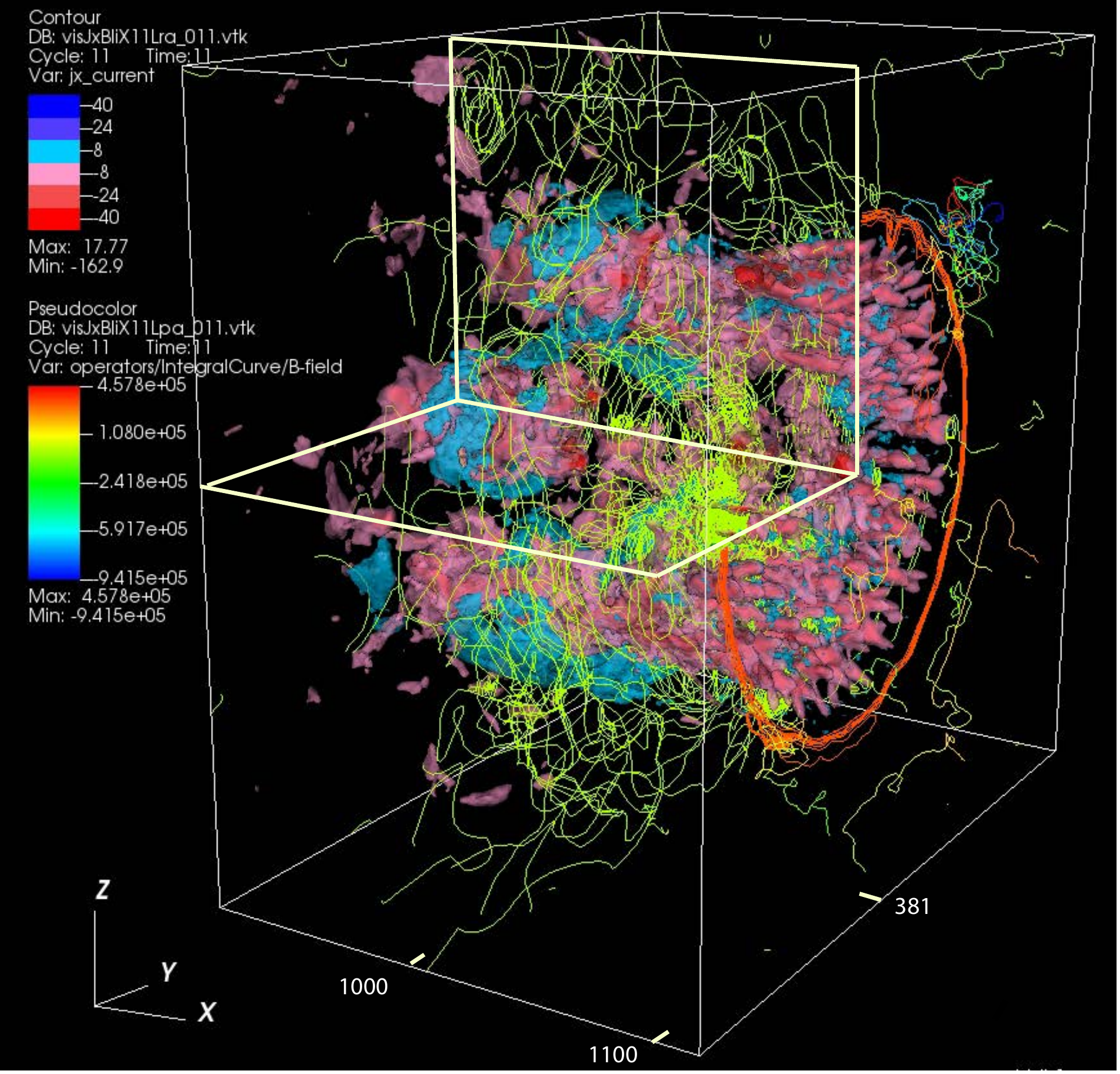}

\vspace{0.1cm}
\caption{Isosurface of the $x$-component of the electron current density, $J_{\rm x}$, with the magnetic field lines \rev{(yellow)} in a rectangular  
section of the simulation grid  ($920 < x/\Delta < 1120; 231 < y/\Delta, z/\Delta < 531$)  
at time $t =1000\, \omega_{\rm pe}^{-1} $. To illustrate the \mpo{distribution of} $J_{\rm x}$ inside the jet, a quadrant of the regions is clipped at the center of \rev{the} jet 
in the $x-z$ plane ($381 < z/\Delta < 531$) and in the $x-y$ plane ($231 < y/\Delta < 381$). The jet front is located at $x/\Delta=1100$.}
\label{3DJX}
\end{figure}

\begin{figure}
\hspace{7.30cm} (a) 

\includegraphics[scale=0.4,angle=0]{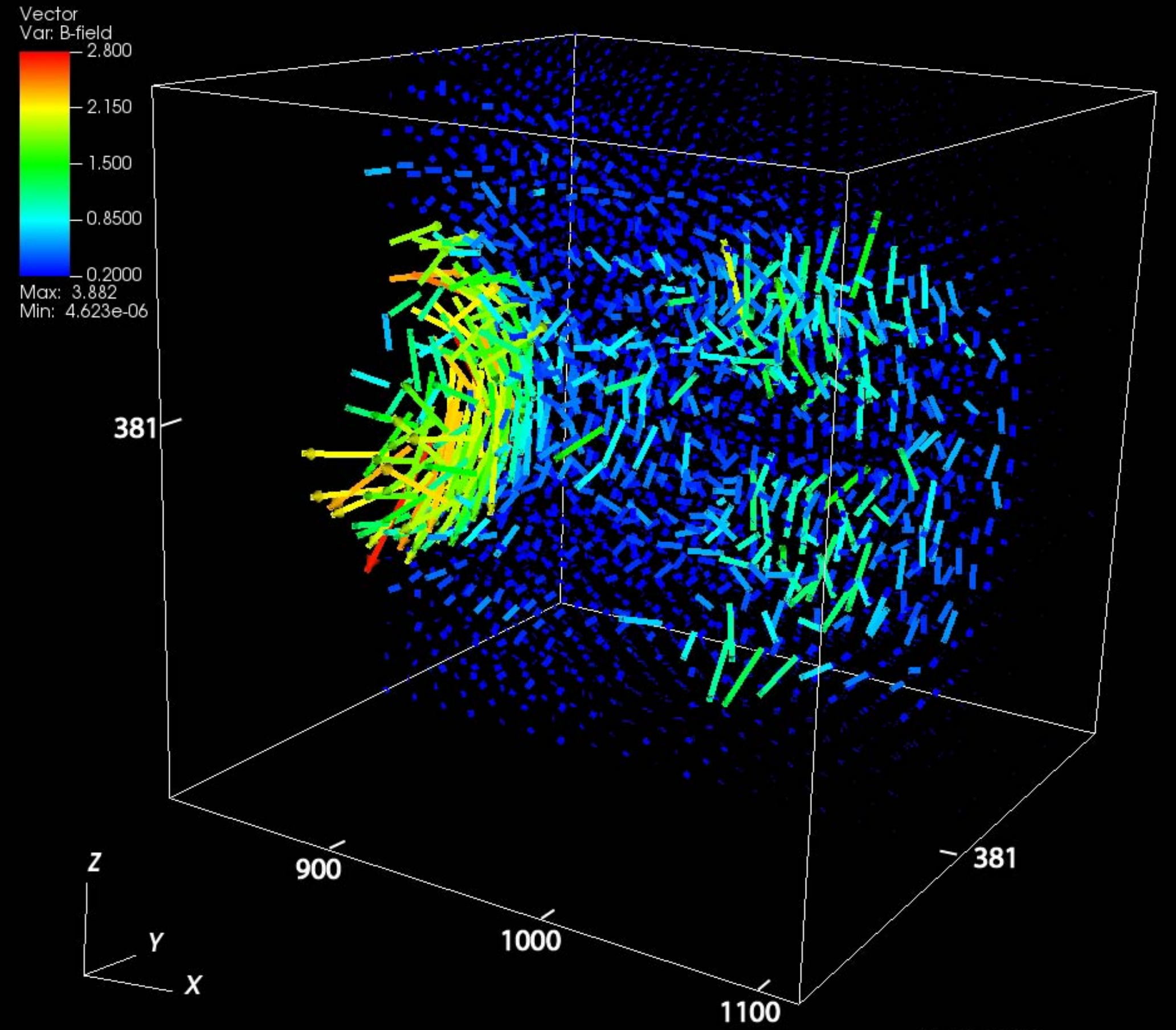}

\vspace{0.1cm}

\hspace{7.30cm} (b) 

\includegraphics[scale=0.4,angle=0]{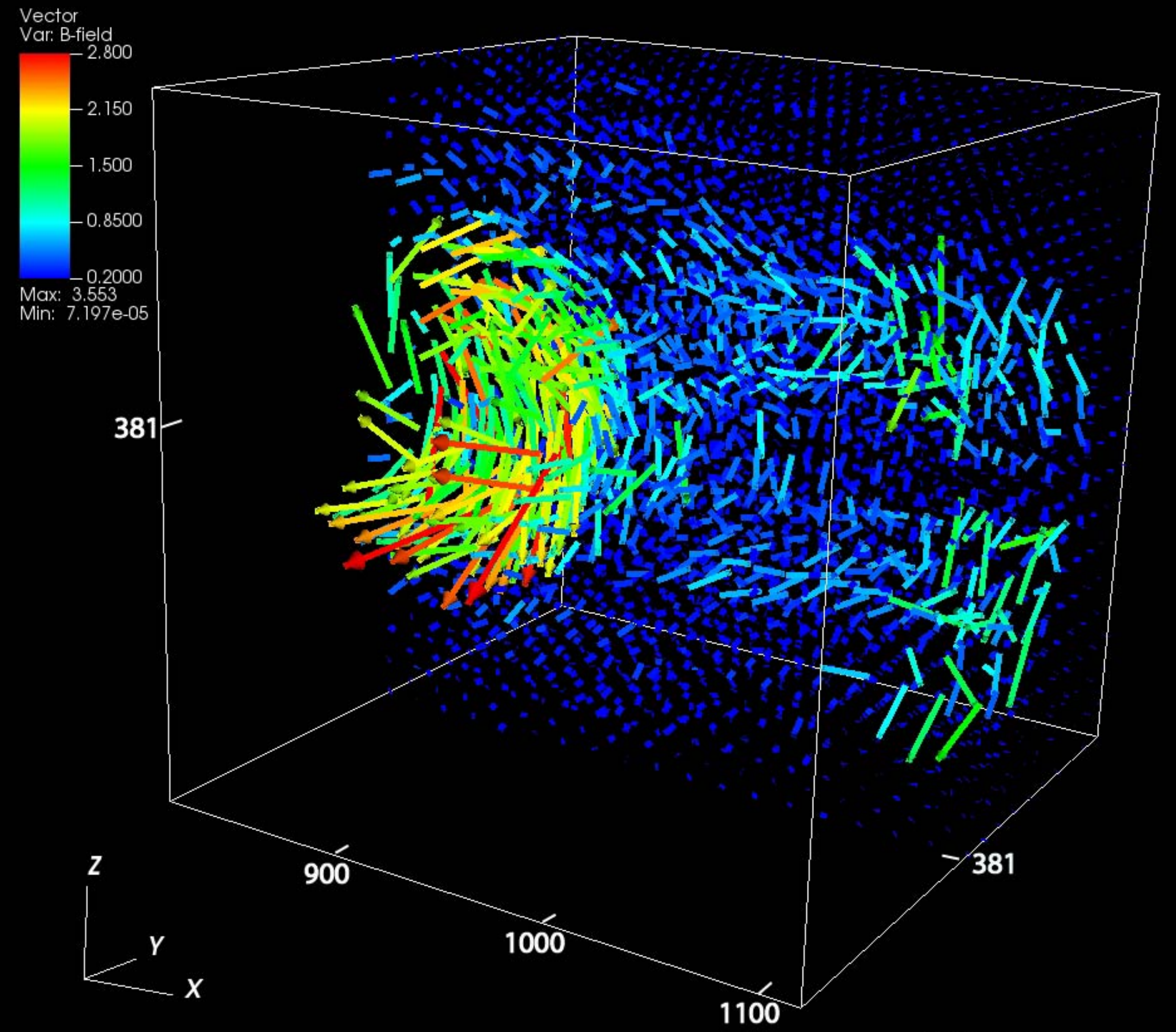}

\vspace{0.1cm}
\caption{Magnetic-field vectors within a cubic section of the simulation grid  ($820 < x/\Delta < 1120; 231 < y/\Delta, z/\Delta < 531$)  
at time $t =900\, \omega_{\rm pe}^{-1} $ (a) and at $t =1000\, \omega_{\rm pe}^{-1}$ (b). To illustrate the magnetic field inside the jet, the plots show \mpo{the rear half of the jet with cut} in the $x-z$ plane ($381 < y/\Delta < 531$).}
\label{3DBV}
\end{figure}

\mpo{In Figure \ref{3DJX} we show three-dimensional isosurfaces of the $x$-component of the electron current density at the jet head together with magnetic field lines plotted in yellow}. Two rectangles indicate the visible area in the jet. Near the jet head, the current filaments generated by the WI are \mpo{evident}. \rev{This result is similar to that obtained by \cite{Ardaneh16}  (their Fig. 3), \mpo{who investigate the structure of the head of an electron-ion jet in} slab geometry with $m_{\rm i}/m_{\rm e}=16$. They \mpo{demonstrated the acceleration of jet electrons in the linear stage of the instability}. Figure \ref{3DJX} shows the merging of current components (the front of the nonlinear region) behind the current filaments, where some jet electrons are slightly accelerated, as we shall show in Fig. \ref{dpx-vx} below.} More complicated structures are seen near the jet boundary, \mpo{that we shall investigate in more detail below.} 

The three-dimensional morphology of the jet's magnetic field is shown in Figure~\ref{3DBV} where we plot magnetic-field vectors 
at $t =900\, \omega_{\rm pe}^{-1}$  (Fig.~\ref{3DBV}a) and  $t=1000\, \omega_{\rm pe}^{-1} $ (Figs.~\ref{3DBV}a and \ref{3DBV}b). The regions 
displayed ($820 < x/\Delta < 1120; 231 < y/\Delta, z/\Delta < 531$) are 
indicated by the red dashed rectangle in Figure~\ref{JxB}b. The plot is clipped at the center of the jet at $y/\Delta =381$. 
One can see that the edge of the helical magnetic field in the jet moves from $x/\Delta =780$  at $t =900\, \omega_{\rm pe}^{-1}$ to 
$x/\Delta = 830$ at  $t =1000\,\omega_{\rm pe}^{-1}$, which is much slower than the jet speed
\rev{(if moving with the jet velocity, the front at $t =1000\, \omega_{\rm pe}^{-1}$ should be located at $x/\Delta\simeq 880$).}
This seems to indicate that the front 
edge of the helical magnetic field is peeled off as the jet propagates. This may indicate that the helical magnetic field is braided by kinetic 
instabilities and subsequently becomes untangled as discussed in \cite{Blandford17}. 
\rev{The untangling of helical magnetic field results from magnetic-reconnection-like phenomena, that push the helical magnetic field \mpo{away from} the center of the jet at the forward position.} Two smaller magnetic islands can be identified in the jet shown in Figure~\ref{3DBV} (half of the jet is shown).
The \rev{supplemental movie}\footnote{\mpo{dBtotByz11MF\_011.mp4: the total magnetic fields in the $y-z$-plane; $280 <y/\Delta<480, 280<z/\Delta<480$}} shows how the helical magnetic field is untangled and magnetic islands evolve along the jet at $t =1000\, \omega_{\rm pe}^{-1}$. For example at $x/\Delta =950$ the centers of magnetic islands are located around $(y/\Delta, z/\Delta)=$ (435, 420) (upper), (440, 318) (lower) (visible in Fig.~\ref{3DBV}b), (320,350), and (280, 400) (not visible).

Figure \ref{dpx-vx} shows the phase-space distribution of jet (red) and ambient (blue) electrons  at $t =900\, \omega_{\rm pe}^{-1}$ and at $t =1000\, \omega_{\rm pe}^{-1}$.
\mpo{Both groups of} electrons are accelerated at several locations that coincide with jet recollimation regions (compare Fig.~\ref{JxB}a-b). 
In particular, electrons are accelerated at $x/\Delta =780$, where the helical 
magnetic field disappears, as shown in Figure~\ref{3DBV}a.  The disappearance of helical magnetic field near $x =830 \Delta$ als coincides with 
the acceleration of electrons. \mpo{Ambient electrons entrained in the relativistic jet plasma} are also strongly accelerated.

\begin{figure}


\includegraphics[scale=0.6,angle=0]{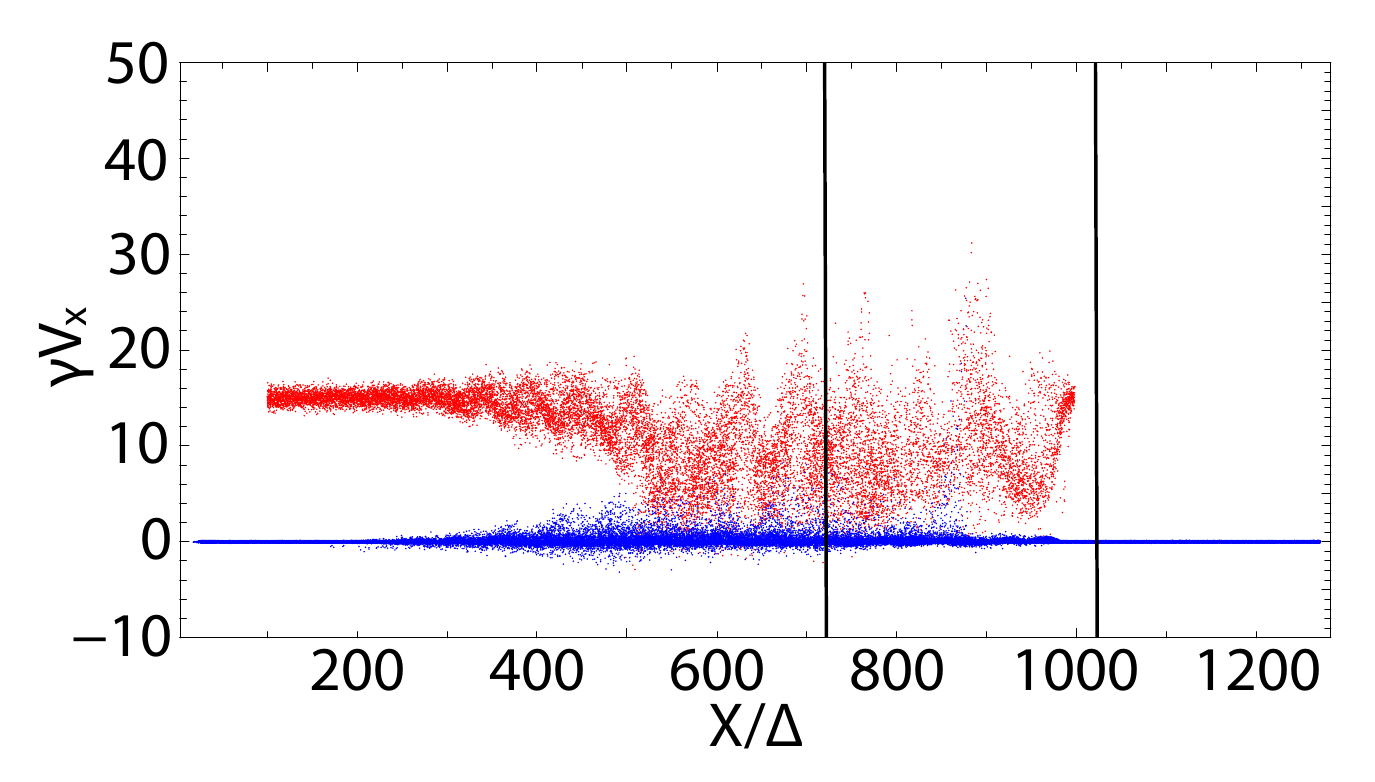}



\includegraphics[scale=0.6,angle=0]{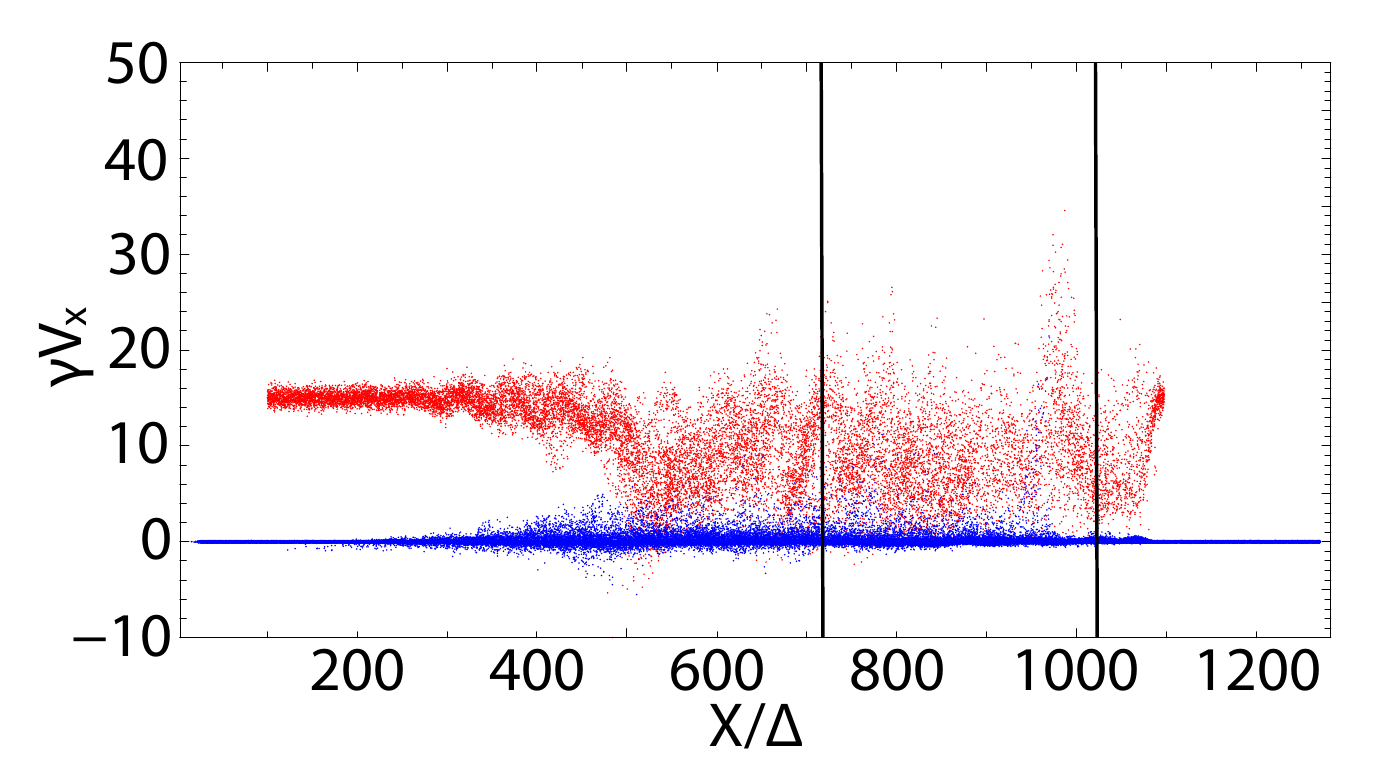}

\vspace{-0.12cm}
\caption{Phase-space ($x-\gamma V_{\rm x}$) distribution of jet (red) and ambient (blue) electrons at $t =900\, \omega_{\rm pe}^{-1} $ (top) and $t =1000\, \omega_{\rm pe}^{-1} $ (bottom). The two vertical lines show the regions \mpo{for which Fig.~\ref{3DBV} displays the 3D magnetic-field vectors.}}
\label{dpx-vx}
\end{figure}

\begin{figure}


\includegraphics[scale=0.6,angle=0]{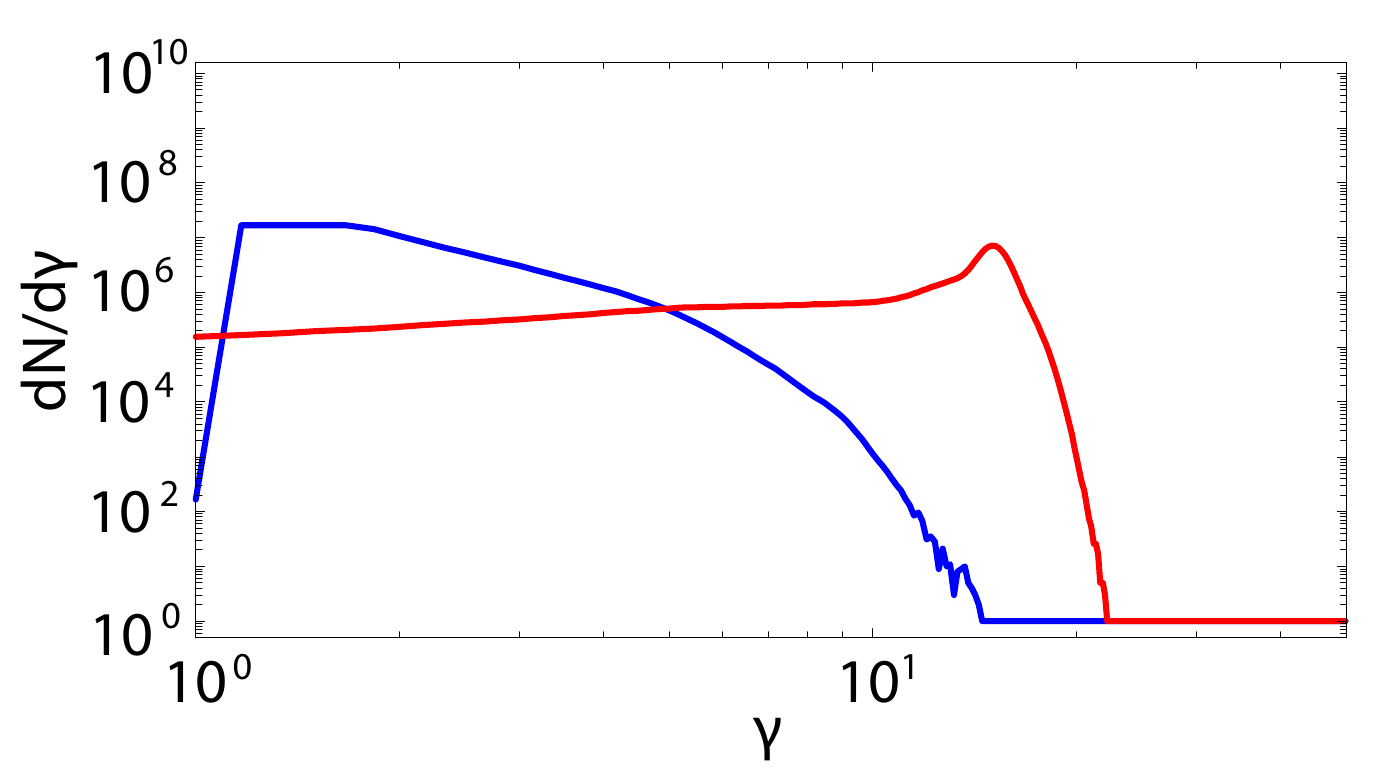}

%

\includegraphics[scale=0.6,angle=0]{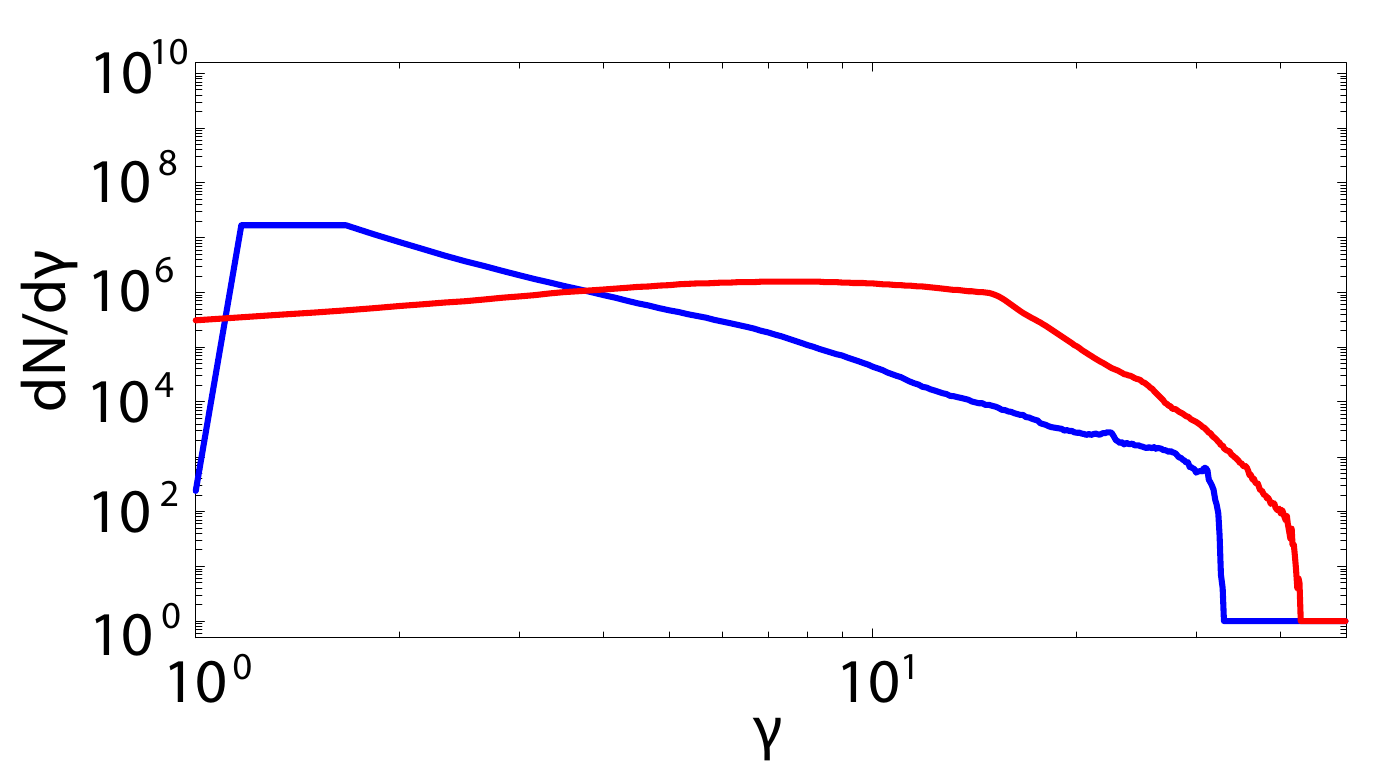}

\vspace{-0.2cm}
\caption{\mpo{Energy distributions of electrons in the region $x/\Delta < 600$ (top panel) and $x/\Delta > 600$ (bottom panel),} corresponding to Fig. \ref{dpx-vx} at 
$t =1000\, \omega_{\rm pe}^{-1}$. The red and blue lines show the jet electron and ambient electron velocity distributions, respectively. }
\label{el-accel}
\end{figure}

\mpo{To examine how electrons are accelerated in the jet, we plot in Fig. \ref{el-accel} their velocity distribution in the linear and the nonlinear stage. The jet electrons initially have a Lorentz factor $\gamma \simeq 15$. The distribution (in red) quickly widens, and some jet electrons are accelerated up to $\gamma = 40$. Surprisingly, the ambient electrons can also be}
accelerated up to $\gamma = 30$.

\begin{figure}


\includegraphics[scale=0.5,angle=0]{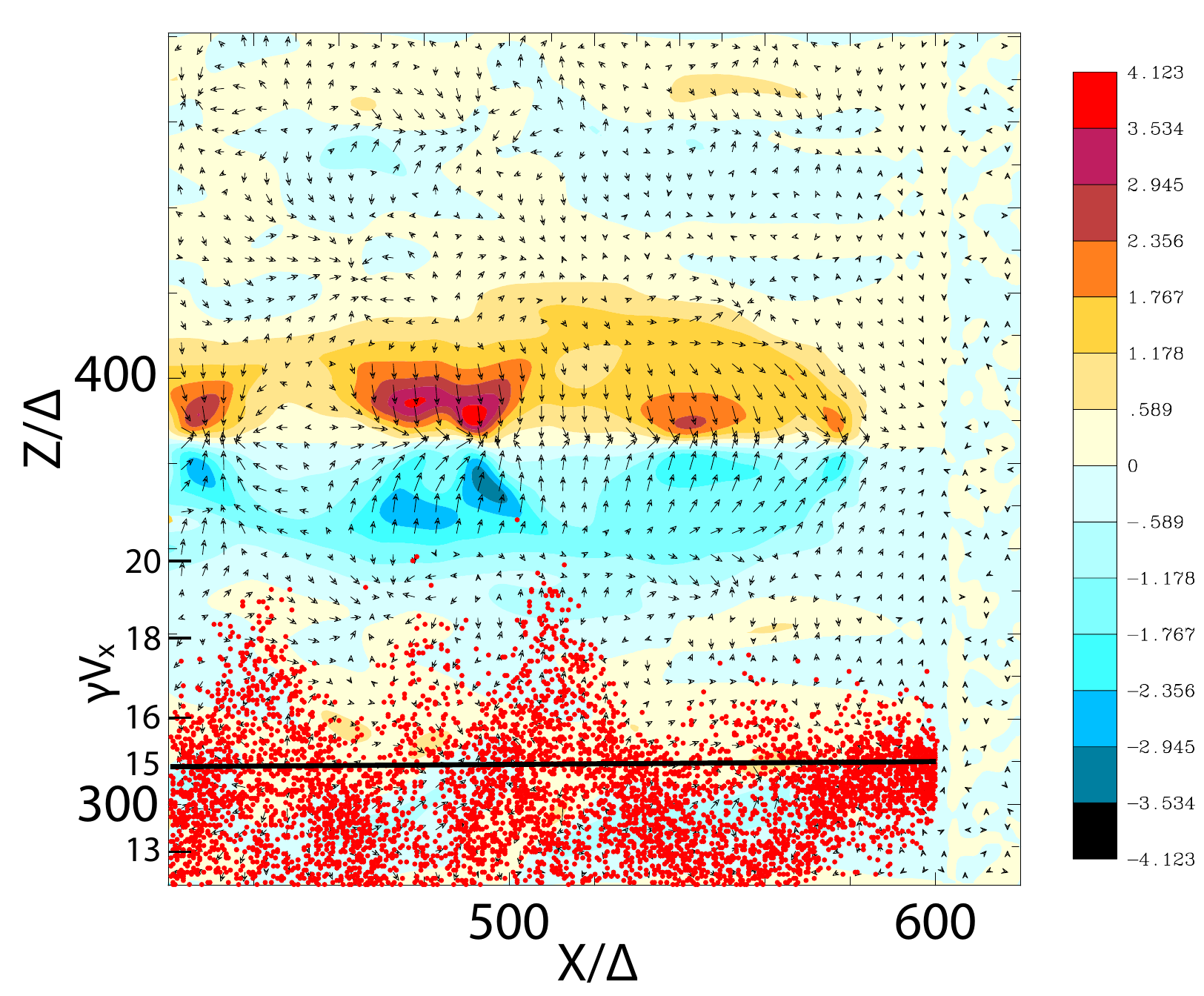}

\vspace{0.1cm}

 
\includegraphics[scale=0.5,angle=0]{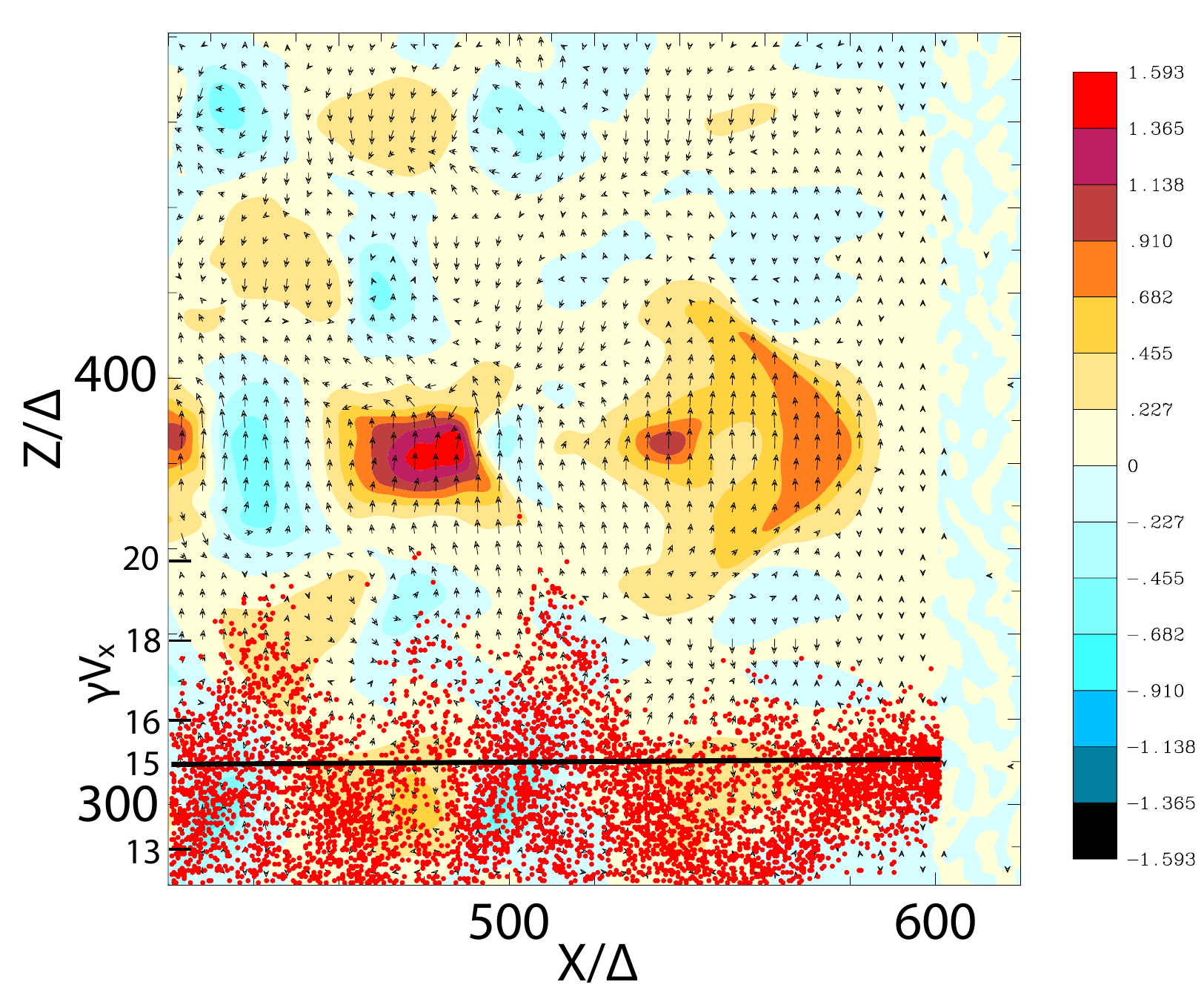}

\vspace{-0.1cm}
\caption{\mpo{Top panel: Contours of $B_{\rm y}$  with arrows of $E_{\rm x, z}$. Bottom:  $E_{\rm x}$ with arrows indicating $B_{\rm x, z}$. The $x-\gamma V_{\rm x}$ distribution of jet 
electrons at $t =500\, \omega_{\rm pe}^{-1}$ (a) is overplotted in red. Some jet electrons are accelerated up to $v_{\rm x}\gamma=20$ 
above the initial $v_{\rm x}\gamma=15$ 
(the black line) in the phase space. The scale for the phase-space is indicated with small digits at} the left axis.}
\label{By-Ex}
\end{figure}

\mpo{Figure~\ref{dpx-vx}  suggests that the groups of accelerated jet electrons visible at $t=900\,\omega_{\rm pe}^{-1}$ moved upward to larger 
$x$ at $t=1000\,\omega_{\rm pe}^{-1}$}. These electrons were accelerated at an earlier time. 
\mpo{To investigate the process of their acceleration}, in Figure~\ref{By-Ex} 
we show the correlation between electron acceleration and structures of the electromagnetic field excited by instabilities 
at $t =500\, \omega_{\rm pe}^{-1}$. 


The MI is the dominant instability in the azimuthal direction. \mpo{Figure \ref{By-Ex} shows that it pinches \rev{the} jet and generates
recollimation shocks. Certainly, kKHI is also excited, \rev{perturbs} 
the jet, and aids in the creation of} 
three recollimation shocks.  \mpo{Figure \ref{By-Ex} features three regions with strong magnetic-field component $B_{\rm y}$, at which we can observe also a positive electric-field component $E_{\rm x}$.} These structures
are similar to recollimation shocks in RMHD simulations (Mizuno et al. 2014). \mpo{The arrows ($B_{\rm x}, B_{\rm z}$)in Figure \ref{By-Ex} indicate magnetic-field reversal at 
$(x/\Delta, z/\Delta) = (440, 390)$. Reconnection in a 3D system is not easily identified in a 2D display, but the evidence in this figure and other diagnostics let's us believe} \rev{that} reconnection takes place. 
The sequence of evolution is \rev{the} following: kKHI and MI (WI) grow, and recollimation shocks are generated. At the same time reconnection occurs. \rev{The} negative quasi-static electric field $E_{\rm x}$ generated by the recollimation shocks accelerates jet electrons. \rev{Electrons are also}
decelerated where a strong positive  $E_{\rm x}$ exists. \rev{We note} that the recollimation \rev{shocks are} not as strong as \rev{shocks generated by the WI}
in slab geometry \citep[compare, e.g.,][]{Ardaneh16}.

Note that 
electrons can be 
further accelerated to higher Lorentz factors on account of turbulent acceleration, as in kinetic simulations of driven 
magnetized turbulence \citep[e.g.,][]{Comisso18,Zhdankin18}. In these simulations turbulent magnetic fluctuations are externally 
forced in the simulation system, and so the energy source for turbulence is not self-consistent. In contrast to the driven turbulence, 
in our simulations turbulent magnetic field (multiple magnetic-field islands) are self-consistently generated in the relativistic jet 
through the untangling of the helical magnetic field.
Particles can be directly accelerated in the reconnection regions and also through interactions with the magnetic islands \mpo{that are visible in, e.g.,} Figure~\ref{JxB}. The particle acceleration processes in turbulent magnetic reconnection have also been investigated by, e.g., \citet[][]{kowal11,kowal12,lazarian2016}.

\section{Summary} \label{sec:Summ}

We performed large-scale three-dimensional PIC simulations of an electron-proton jet containing helical magnetic fields. \mpo{The aim is to investigate 
in a systematic way how the helical magnetic field in jets evolves and accelerates electrons. 
The electron-proton jet undergoes kinetic instabilities, the most dominant of which is the MI. It pinches the jet and generates recollimation shocks. The kKHI also operates and contributes to the creation of recollimation shocks. We observed a number of instances of strong transverse magnetic field, $B_{\rm y}$, coinciding with significant jet-parallel electric field, $E_{\rm x}$, that are similar to the structure of recollimation shocks in RMHD simulations (Mizuno et al. 2014). 
Using as tracer magnetic-field reversals and other diagnostics, we found regions in which magnetic reconnection takes place. }

\mpo{It appears that first kKHI and MI (WI) grow, then recollimation shocks are generated, and reconnection operates. The recollimation shocks feature a quasi-static electric field, $E_{\rm x}$, that depending on its sign accelerates or decelerates jet electrons.}
In the nonlinear stage the toroidal magnetic fields are untangled.
The disappearance of helical magnetic fields 
generates magnetic field islands 
\rev{that interact with each other and}
accelerate electrons.
\cite{kowal11,kowal12,lazarian2016} 
\rev{investigated} the particle acceleration process in turbulent reconnection theoretically and by numerical experiments. They \mpo{demonstrated} an exponential growth in the energy of the accelerated particles and also the development of a power-law tail in the particle spectrum, \mpo{both of} which are signatures of stochastic Fermi acceleration by reconnection. Figure \ref{el-accel} shows
a power-law tail \mpo{and hence is consistent with this scenario, but further investigation are needed to confirm stochastic Fermi acceleration.}

Recently, \citet{Pino18,Kadowaki18,Kadowaki19, singh2016} \mpo{used relativistic MHD simulations with test particles to investigate acceleration through turbulent magnetic reconnection driven by \rev{KI}. 
\rev{Using PIC simulations}
\cite{alves18,alves19} found that the formation of highly tangled magnetic fields and a large-scale inductive electric field in kink-unstable jets accelerates particles through curvature drift. 
%
%
There is no bulk flow in their simulations}, \rev{and} therefore velocity-shear instabilities such as kKHI and MI \rev{are} not excited, \rev{and KI occurs instead}.
\mpo{Since the magnetization in their simulations is very large, $\sigma>1$,  the toroidal magnetic field is dominant and the KI is triggered,}
which tangles the helical magnetic field and consequently may result in
reconnection, \mpo{and not necessarily only} curvature drift acceleration. 
\mpo{We should add that the jet density profile assumed in \citet{alves18}, $n=n_0 +(n_{\rm c} - n_0)/\cosh^{2}(2r/R_{\rm c})$, differs from the top hat profile that we employ. Our choice may be more conducive to the  kKHI, MI, and the Weibel modes. }

Since kinetic KHI and MI grow faster than current-driven kink instabilities due to strong velocity shear and moderate magnetization,
our simulation does not show a kink-like instability as seen in the simulations of \cite{alves18}. In our simulation the 
particles are accelerated by the turbulent magnetic reconnection which is initiated by the growth of kinetic instabilities
such as kKHI and MI.  

\mpo{We investigated in a self-consistent fashion one of the possible mechanisms of electron acceleration in relativistic jets containing helical magnetic field. We demonstrated that as the electron-proton jet evolves, the helical magnetic field becomes untangled and locally reconnects. Electrons are accelerated within the resulting turbulent magnetic field. More simulations with larger 
jet radii are required, in which we vary the magnetic-field structures, for example the characteristic scale, $a$, and the pitch profile 
parameter, $\alpha$. Moreover, a variation of the density profile and the magnetization of the jet are required 
and will be the subject of future studies.}

\section*{Acknowledgements}

We appreciate Christoph K\"ohn's critical reading and fruitful suggestions which improved the contents of this report.
This work is supported by the NASA-NNX12AH06G, NNX13AP-21G, and NNX13AP14G grants. The recent
work is also provided by the NASA through Chandra Award Number GO7-18118X (PI: Ming Sun at UAH) issued by the Chandra X-ray Center, which is operated by the SAO for and on behalf of the NASA under contract NAS8-03060.
The work of J.N. and O.K. has been supported by Narodowe Centrum Nauki through research project DEC-2013/10/E/ST9/00662. Y.M. is supported by the ERC Synergy
Grant ``BlackHoleCam: Imaging the Event Horizon of Black Holes'' (Grant No. 610058). The work of I.D. has been supported by the NUCLEU project. Simulations were performed using Pleiades and Endeavor facilities at NASA Advanced Supercomputing (NAS: s2004), and using Comet at The San Diego Supercomputer Center (SDSC), and Bridges at The Pittsburgh Supercomputing Center, which are supported by the NSF.












\bsp	
\label{lastpage}
\end{document}